\documentclass[fleqn,10pt]{wlscirep}

\usepackage[]{graphicx}
\DeclareGraphicsExtensions{.pdf,.png,.jpg}
\setkeys{Gin}{draft=false}

\usepackage{gensymb}
\usepackage{amsmath}
\usepackage{amssymb}

\title{Effects of Pore-Scale Disorder on Fluid Displacement in Partially-Wettable Porous Media}

\author[1,*]{Ran Holtzman}
\affil[1]{Department of Soil and Water Sciences, The Hebrew University of Jerusalem, Rehovot, Israel}
\affil[*]{holtzman.ran@mail.huji.ac.il}

\usepackage{setspace}
\begin{document}

\begin{abstract}
We present a systematic, quantitative assessment of the impact of pore size disorder and its interplay with flow rates and wettability on immiscible displacement of a viscous fluid. 
Pore-scale simulations and micromodel experiments show that reducing disorder increases the displacement efficiency and compactness, minimizing the fluid-fluid interfacial area, through (i) trapping at low rates and (ii) viscous fingering at high rates. 
Increasing the wetting angle suppresses both trapping and fingering, hence reducing the sensitivity of the displacement to the underlying disorder. 
A modified capillary number $\mathrm{Ca^*}$ that includes the impact of disorder $\lambda$ on viscous forces (through pore connectivity) is direct related to $\lambda$, in par with previous works.  
Our findings bear important consequences on sweep efficiency and fluid mixing and reactions, which are key in applications such as microfluidics to carbon geosequestration, energy recovery, and soil aeration and remediation.
\end{abstract}

\flushbottom
\maketitle
\thispagestyle{empty}


\section*{Introduction}

In many of the challenges we face today as geoscientists, in particular in the context of water and energy resources, fluid invasion into a porous soil or sediment is a key process. Examples include hydrocarbon migration and recovery, methane venting from hydrate-bearing sediments, drying and wetting of soils, and carbon geosequestration~\cite{sahimi-ftpm}. Fluid displacement has also gained attention as a scientific paradigm for problems where competitive domain growth and nonlinear interface dynamics are influenced by spatial heterogeneity of the system properties, such as magnetic domains and biological films~\cite{Pelce2004}. 

The complex interplay between capillary, viscous, and gravitational forces, wettability effects, and the underlying \textit{heterogenous} pore geometry, leads to ramified, preferential flow paths or ``fingering''. Predicting the emergent patterns is challenging, because of the sensitivity to pore-scale details and the large number of coupled mechanisms and governing parameters which vary over a wide range of values and scales~\cite{Meakin2009,Bultreys2016}. 
Fingering leaves pockets of the original, defending fluid, increasing the fluid-fluid interfacial area, with important consequences on efficiency of processes relying on fluid withdrawal such as enhanced hydrocarbon recovery~\cite{LakeOil2014}, or on fluid mixing and reaction such as subsurface remediation~\cite{Nadim2000} or CO$_2$ sequestration~\cite{Matter2016}.


The practical importance of fluid displacement motivated intensive research, that identified several important mechanisms governing the displacement efficiency. 
In disordered media, slow displacement of a wetting fluid (by a less wetting fluid, termed ``drainage'') leads to a patchy pattern with multiple islands of defending fluid, called capillary fingering (CF). 
The process of forming these islands is called trapping, as lack of fluid continuity between them inhibits their mobilization.
Increasing the flow rate (hence the strength of viscous forces relative to capillary ones, characterized by the capillary number $\mathrm{Ca} = \mu_{d}v/\gamma$) modifies the pattern depending on the fluids viscosity ratio, $M= \mu_d / \mu_i$, towards (i) viscous fingering (VF) for an unfavorable ratio, $M  > 1$, or (ii) compact (CO) for a favorable ratio ($M < 1$) \cite{lenormandtouboul88,lenormand90-liquids}. 
Here $\mu$ is the viscosity, $v$ is the characteristic fluid velocity, and $\gamma$ is the interfacial tension, where subscripts $d$ and $i$ denote defending and invading, respectively.

Another important property is wettability, namely the relative affinity of the fluids to the solid. Increasing affinity of the invading fluid (from drainage to imbibition, when the invading fluid is more wetting) smoothes the invasion front, even at high, unfavorable $M$~\cite{stokesweitz86}. At low rates, increasing wettability leads to a transition between CF and CO, whereas increasing $\mathrm{Ca}$ enhances viscous instabilities promoting VF, regardless of wettability~\cite{Trojer_PRAP2015,Holtzman2015}. 
%
We note that the frequent use of the term ``(in)stability'' to describe the pattern, and the interchange of ``stability'' with ``efficiency'' or ``compactness''. To avoid confusion between \textit{patterns} (e.g. VF) and underlying \textit{mechanisms} (e.g. hydrodynamic instabilities), we use ``compact'', ``efficient'' or ``stable pattern''.


%

%



Experimental and computational capabilities limit the number of mechanisms that can be modeled. Here, we focus on the following subset: capillarity, viscosity, wettability, and their coupling with random disorder in pore and/or particle sizes. 
%
%
We exclude here gravity and buoyancy effects. We also focus on \textit{size} disorder alone, excluding other types of heterogeneity including in \textit{wetting properties}~\cite{Murison2014} or with spatial \textit{correlations} such as layers or dual porosity media~\cite{Zhang2011,AkhlaghiAmiri2014}.
%

\subsection*{Current understanding of disorder effects}

Disorder in pore sizes is inherent in natural media such as soils and rocks~\cite{Bultreys2016}, and, to some extent, even in engineered devices due to limited fabrication resolution~\cite{Stone_microfluidics2004}. 
In some industrial applications, where homogenous fluid distribution and control over resulting patterns are key, low disorder is crucial and highly desirable~\cite{Kim2015}.
We note that disorder as well as preferential pathways and heterogenous fluid distributions can be advantageous in applications where fluid mixing and reactions~\cite{Nadim2000,Matter2016} are needed, and undesirable when efficient sweep of hydrocarbons~\cite{LakeOil2014} or contaminants~\cite{Nadim2000} are crucial. 
As summarized below, while the importance of disorder is widely-recognized, a systematic exploration including quantitative analysis of samples of different, known disorder at variable wettability and rates is lacking.

 
%


The effect of disorder on the invasion morphology when \textit{either} capillary or viscous forces are dominant seems to be univocally accepted. 
For slow drainage, high disorder was shown to promote trapping, leading to a transition from CO to CF \cite{yortsosxu97, Cieplak1988,Cieplak1990, toussaintlovoll05, holtzmanjuanes10-fingfrac , Xu2014, Liu2015}. For slow imbibition, quasi-static simulations show a transition from compact to faceted (FA) growth as disorder is decreased~\cite{Cieplak1990, MartysPRB1991}. Ordered FA patterns were also observed experimentally, for imbibition in ordered samples (uniform pore sizes) at favorable $M$~\cite{lenormand90-liquids}. 
%
%
An interesting correlation between fluid invasion and magnetic systems was drawn regarding the interplay between disorder and the smoothing (``stabilizing'') action of cooperative events~\cite{Martys1991}. 
%
At the other extremity of high $\mathrm{Ca}$ (VF regime), decreasing disorder produces snowflake-like patterns~\cite{chenwilkinson85,holtzmanjuanes10-fingfrac} with fewer, less tortuous fingers (``ordered dendritic'', OD). 
We stress that FA and OD are only obtained for samples with nearly-uniform pore sizes, here resulting from uniformly sized particles ordered on a lattice. Such patterns would not appear in granular materials (including monosized bead packs) due to the large variability in apertures provided by \textit{packing} disorder. 
%
%

At intermediate $\mathrm{Ca}$ where \textit{both} viscosity and capillarity are important, the qualitative paradigm (``conventional wisdom''~\cite{Liu2015}) is that lowering disorder enhances the displacement efficiency~\cite{yortsosxu97, toussaintlovoll05, holtzmanjuanes10-fingfrac,Xu2014,Liu2015}.
%
Few works derived a modified capillary number, $\mathrm{Ca}^*$, which includes the impact of disorder via the width of the capillary thresholds, neglecting its effect on the viscous forces.
Quantitatively, this implies an inverse relationship between $\mathrm{Ca}^*$ and disorder, namely that increasing disorder (at a given rate, $\mathrm{Ca}$) produces a more CF-like pattern, increasing the transitional $\mathrm{Ca}$ between CF to VF, $\mathrm{Ca_{CF/VF}}$ \cite{yortsosxu97, toussaintlovoll05, holtzmanjuanes10-fingfrac,Xu2014,Liu2015}. 
We point to a subtle yet crucial inconsistency in the above: if increasing disorder increases $\mathrm{Ca_{CF/VF}}$, promoting a more CF-like pattern, sweep efficiency should increase (since VF is much less efficient that CF \cite{yortsosxu97, toussaintlovoll05, holtzmanjuanes10-fingfrac,Toussaint2012,Xu2014,Liu2015}), contrasting the observations stated above.


Here, we address this problemtic paradigm via a combination of pore-scale simulations, micromodel experiments and scaling analysis. 
%
To the best of our knowledge, our study is the first systematic, quantitative investigation of the impact of disorder and its coupling with rates and wettability. 
%
%
Our findings improve our understanding, generalizing the phase diagram of fluid invasion including rates, viscosity ratio, buoyancy and wettability effects \cite{lenormandtouboul88,Toussaint2012, Trojer_PRAP2015,Holtzman2015} by adding an additional axis--disorder.




\section*{Results} 

\subsection*{Problem setting and methodology}   

\begin{figure}[h]
\centering
\noindent\includegraphics[width=.85\columnwidth]{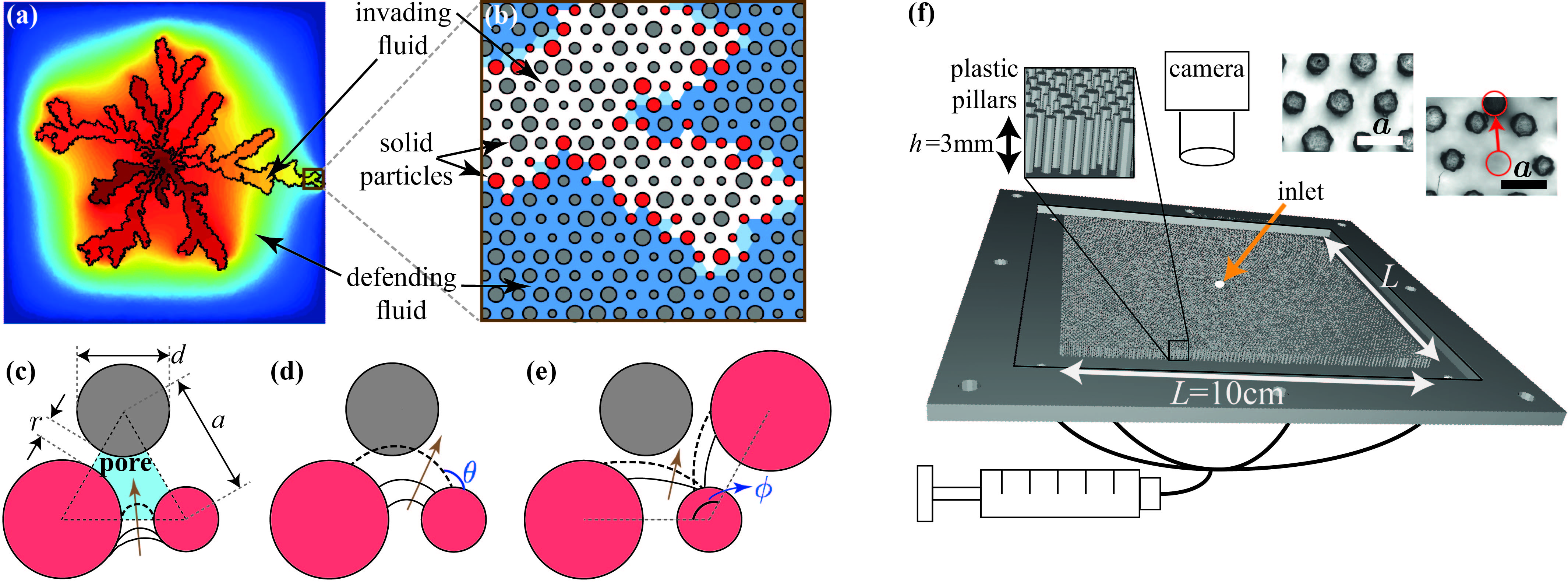}
\caption{Our porous medium is made of variably-sized cylindrical pillars placed on a triangular lattice. 
(a) Numerically, we track the fluid-fluid interface (black line) and fluid pressures (increasing from blue to red).
(b) Zoom in showing the lattice (spacing $a$) of particles (diameter $d$) and pores interconnected by throats (of width $r$, see panel c). The fluid-fluid interface is represented as a sequence of circular menisci, touching particles at contact angle $\theta$, with radius of curvature $R$ set by the local capillary pressure. Menisci can be destabilized by: (c) {burst}; (d) {touch}; or (e) overlap. Brown arrows indicate direction of advancement, destabilized arc in dash (From Holtzman and Segre~\cite{Holtzman2015}; Copyright 2015 by the American Physical Society). 
(f) Experimental setup: A micromodel is initially saturated with glycerol, which is withdrawn by a syringe pump from the cell's perimeter (outlet), allowing air to invade at its center (inlet). 
Time-lapse images provide the pattern evolution. Limited fabrication resolution results in pillars with rough, non-round edges, and, in a few places, misplaced (inset, in red; $a=1$ {mm}). 
\label{fig:model_and_expt_schematics}}
\end{figure}

We consider radial displacement of a viscous fluid by an inviscid fluid (high, unfavorable ratio $M$) at a fixed rate $Q$ in samples of different disorder in particle sizes. 
Details of the simulations and experiments are provided in Methods and Supplementary Material. Briefly, our 2-D medium is made of ``particles'' (cylindrical pillars) of diameters $d$
drawn from a uniform distribution of width $\lambda$, $d\in[1-\lambda,1+\lambda]\overline{d}$, providing well-defined pores and throats (of size $r$, with a triangular distribution), see Fig.~\ref{fig:model_and_expt_schematics}a--c.
Experimentally, we use a plastic micromodel initially saturated with glycerol, withdrawn from the cell's perimeter (outlet), allowing air to invade at its center. 
Time-lapse photography provides the pattern evolution (Fig.~\ref{fig:model_and_expt_schematics}f). 
The large data set required for the quantitative analysis, including 240 runs on samples of different disorder, at various rates and wetting properties, was generated numerically, with the pore-scale model of Holtzman and Segre~\cite{Holtzman2015}. 

%
%

%

\subsection*{Simulated patterns}   
\label{Simulated_patterns} 

\begin{figure}[h]
\centering
\includegraphics[width=.80\columnwidth]{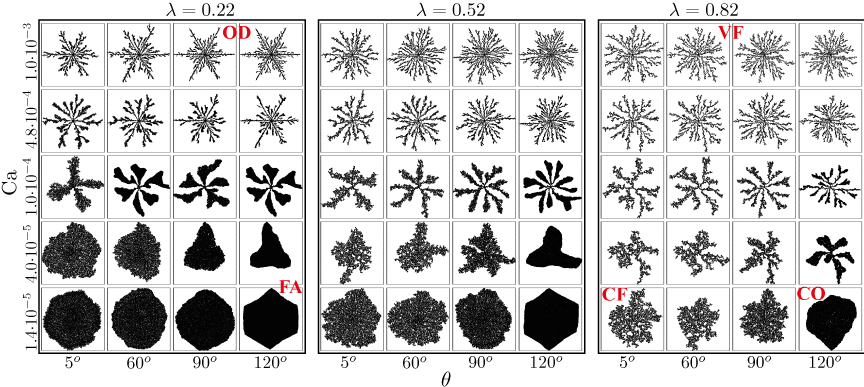}
\caption{Simulated displacement patterns at various flow rates ($\mathrm{Ca}$) and wetting properties ($\theta$), for three degrees of disorder, $\lambda$. Increasing $\lambda$ decreases the displacement efficiency, in a manner that depends on both $\mathrm{Ca}$ and $\theta$: promoting trapping and capillary fingering (CF) at slow drainage (low $\mathrm{Ca}$ and $\theta$), and viscous fingering (VF) at high rates (regardless of $\theta$). Reducing $\lambda$ enhances the effect of the underlying geometry (the ordered lattice), which, together with the stabilizing effect of wettability results in faceted (FA) and ordered dendritic (OD) patterns at low and high $\mathrm{Ca}$, respectively.
Pores filled with invading and defending fluid are marked in black and white, respectively (solid not shown).  Sample size is $L$=$260a$.
\label{fig:patterns}}
\end{figure}

Our simulations capture the plethora of invasion behaviors and emergent patterns as the disorder, rate and wettability are varied. For high disorder ($\lambda$=0.82), we reproduce the transition from CF to VF as the rate increases in drainage~\cite{lenormandtouboul88} (wetting angle $\theta< 90^{\circ}$, $\theta$ measured through the defending fluid, see Fig.~\ref{fig:model_and_expt_schematics}c), and the smoothing (``stabilizing'') effect of increasing $\theta$ towards imbibition ($\theta> 90^{\circ}$), reducing trapping~\cite{Trojer_PRAP2015,Holtzman2015} (Fig.~\ref{fig:patterns}). 
We show that lowering the disorder $\lambda$ enhances the displacement efficiency, by inhibiting trapping and formation of thin, tortuous fingers at low and high rates, respectively.
While in highly-disordered systems the pattern is random (showing no dependence on the underlying structured lattice), reducing $\lambda$ amplifies the impact of the underlying lattice, suppressing both trapping \textit{and} fingering, eventually resulting in the following transitions: (i) from CF to CO in slow drainage \cite{Trojer_PRAP2015}; (ii) from CO to FA in slow imbibition \cite{MartysPRB1991}; and (iii) from VF to OD \cite{chenwilkinson85,holtzmanjuanes10-fingfrac} at high rates. 
We point that compact displacement is obtained despite of the highly unfavorable viscosity ratio, $M = \mu_d / \mu_i \approx 55$.
Reducing $\lambda$ also suppresses the impact of wettability; that is, the competing effects of reducing $\lambda$ and increasing $\theta$ offset each other (Fig.~\ref{fig:patterns}).
%



%

\subsection*{Experimental validation}   
\label{expt_results} 

\begin{figure}[h]
\centering
\includegraphics[width=.6\columnwidth]{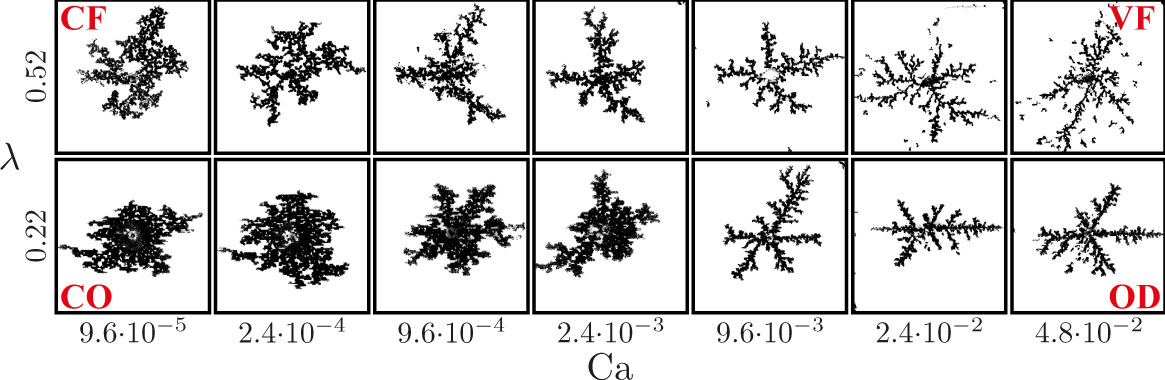}
\caption{Experimental displacement patterns patterns in micromodels with different disorder $\lambda$, showing that decreasing $\lambda$ provides a more ordered pattern, with a transition from CF to CO and from VF to OD at low and high rates, respectively, in agreement with our numerical simulations. 
%
In these difference images, showing changes since air invasion started, air appears black, and solid pillars and glycerol (the defending fluid) are invisible (white).
\label{fig:experiments}}
\end{figure}

To validate our numerical findings, we conducted micromodel experiments with two cells of different disorder ($\lambda$=0.22 and 0.52) of geometry identical to the simulations, displacing glycerol by air ($\mu_d \approx 1$Pa$\cdot$s, $\mu_i = 1.8 \cdot 10^{-5}$Pa$\cdot$s, $\gamma\approx$ 6.3$\cdot$10$^{-2}$N/m, $\theta = 75^{\circ}\pm5^{\circ}$, see Methods).
The experimental patterns support our findings that reducing $\lambda$ enhances the displacement efficiency by (i) suppressing trapping at low $\mathrm{Ca}$, leading to a transition from CF to CO; and (ii) reducing the number of tortuous fingers, increasing their width (from VF to OD) at high $\mathrm{Ca}$ (Fig.~\ref{fig:experiments}). 
The experiments also show the impact of $\lambda$ on the invasion dynamics: (i) the transition in the mode of front propagation from radially-symmetric and continuous (CO) to intermittent and asymmetrical (CF) as $\lambda$ is increased at low rates (Videos 1 and 2 in the the supplementary material); and (ii) radial growth of viscous fingers with vanishing impact of the underlying lattice as $\lambda$ increases (from OD to VF, see Videos 3 and 4) at high rates.

We note that while patterns in simulations and experiments are similar in type (e.g. number and width of fingers or interface roughness), they are not identical. 
Potential causes for this mismatch include (1) manufacturing defects in pillar placement and shape (Fig.~\ref{fig:model_and_expt_schematics}f, insets) and (2) mechanisms not included in our model such as snapoff~\cite{lenormand90-liquids,Joekar-Niasar_CREST_2012} or droplet fragmentation~\cite{Pak2015} in low-porosity rocks, and thin wetting fluid films progressing ahead of the main imbibition front~\cite{Zhao2016}.
Manufacturing errors could greatly affect the pattern due to its sensitivity to small geometrical details; e.g. enlarging a single aperture can allow invasion of multiple pores, otherwise inaccessible. 
In contrast, for the current settings of relatively high viscosity contrast $M$, low contact angle (vs. $150^{\circ}$ in \cite{Zhao2016}), high porosity and small throat to pore size ratio, the aforementioned mechanisms should not be dominant~\cite{lenormand90-liquids,Joekar-Niasar_CREST_2012}.
A more conclusive validation requires improved manufacturing accuracy as well as a more rigorous, quantitative analysis. Such experiments are underway. Modeling mechanisms such as droplet fragmentation or film flow is a challenge which is outside the scope of this paper~\cite{Meakin2009,Joekar-Niasar_CREST_2012,Bultreys2016}, however of great interest for future research. 

\subsection*{Pattern characteristics}   
\label{Pattern_characteristics} 

We characterize the patterns through their finger width $w$, interfacial areas $A\mathrm{_{inter}}$ and $A\mathrm{_{front}}$, and breakthrough saturation $S$. 
The interfacial areas (lengths in 2-D) are computed from the ratio of interfacial and invaded pores, including trapped regions ($A\mathrm{_{inter}}$) or excluding them ($A\mathrm{_{front}}$). The areas are normalized by the invaded volume $V_\mathrm{inv}$ (area in 2-D), and multiplied by the lattice spacing $a$ to make them dimensionless.
The finger width $w$ is evaluated by skeletonizing the pattern with a Voronoi algorithm, and measuring the distance (in lattice units $a$) of the medial line from the interface. 
In evaluating $w$ we include small clusters trapped \textit{within} fingers, unlike some previous studies where these were ignored~\cite{Toussaint2012} (resulting in larger $w$).

\begin{figure}[h]
\centering
\includegraphics[width=.9\columnwidth]{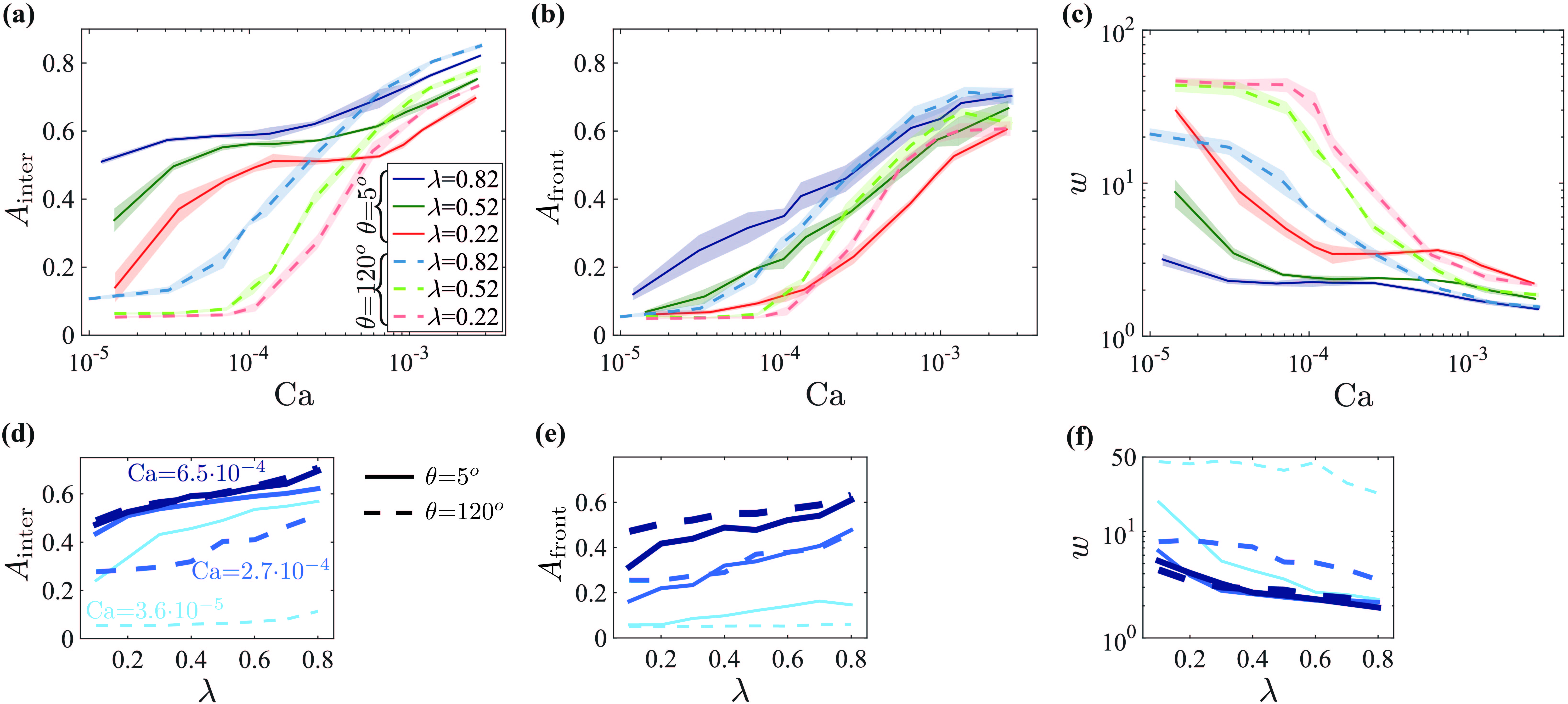}
\caption{Simulated pattern characteristics: (a) fluid-fluid interfacial area $A\mathrm{_{inter}}$; (b) front area (excluding trapped regions) $A\mathrm{_{front}}$; and (c) mean finger width $w$.
The improved displacement efficiency as $\lambda$ is lowered is manifested by a decrease in $A\mathrm{_{inter}}$ and $A\mathrm{_{front}}$, and increase in $w$. Efficiency is also enhanced by decreasing rate ($\mathrm{Ca}$) and increasing wettability ($\theta$).
Slow imbibition ($\mathrm{Ca}$=3.6$\cdot$10$^{-5}$, $\theta$=120$^{\circ}$, light blue in d--f) produces a compact front which is nearly unaffected by disorder. 
Lines and shading in a--c depict the ensemble mean and standard deviation among four realizations (with $L$=$100a$). 
Panels d--f show data from eight samples ($\lambda$) and three rates (darker and thicker lines indicate higher $\mathrm{Ca}$).
\label{fig:Linter_Lfront_w}}
\end{figure}

Our simulations point to the persistent effect (across variable $\mathrm{Ca}$ and $\theta$) of increasing disorder $\lambda$, manifested as an increase in $A\mathrm{_{inter}}$ and $A\mathrm{_{front}}$ (approaching the theoretical limit of 1 for VF at high rates), and a decrease in $w$ (Fig.~\ref{fig:Linter_Lfront_w}a--c). 
The counteracting effects of increasing $\theta$ and lowering $\lambda$ are demonstrated by a nearly-constant, low $A\mathrm{_{inter}}$ and $A\mathrm{_{front}}$ and high $w$ for $\lambda<0.6$, in particular at low $\mathrm{Ca}$ (approaching the limit $a/L$ for CO), see Fig.~\ref{fig:Linter_Lfront_w}d--f.
%
As shown below, this insensitivity is caused by two competing effects of decreasing $\lambda$ in imbibition (high $\theta$): (i) reducing the variation in capillary thresholds, increasing compactness; and (ii) inhibiting cooperative events which smooth the interface, hence decreasing compactness.

\subsection*{Invasion selectivity and cooperativity} 

\begin{figure}
\centering
\includegraphics[width=.8\columnwidth]{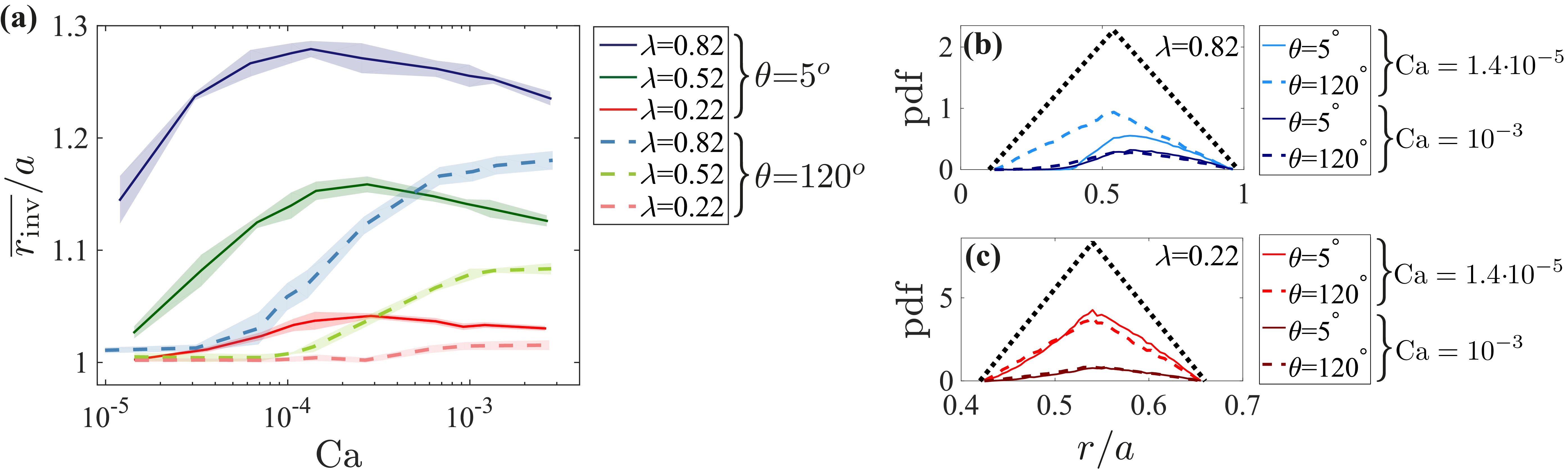}
\caption{Increasing disorder increases the invasion selectivity, namely the mean invaded pore throat size $\overline{r_\mathrm{inv}}$ (panel a) by avoiding small pores (b). At low $\lambda$, the invasion samples nearly equally all throat sizes (c), such that $\overline{r_\mathrm{inv}} \approx \overline{r} =0.54 a$.  
Selectivity is further reduced with the wetting angle $\theta$, due to increasing occurrence of overlaps which are less sensitive to aperture sizes than bursts. 
In contrast, selectivity increases with rates, as screening produces thin fingers which completely bypass low permeability regions.
Lines and shading in (a) are ensemble mean and standard deviation among four realizations. 
Panels b and c show the probability density function (pdf) of \textit{invaded} throat sizes (${r\mathrm{_{inv}}}$, colored lines), and the triangular throat size distribution ($r$, black dots) for two samples, $\lambda$=0.82 (b) and 0.22 (c). 
\label{fig:select}}
\end{figure}

To further explain the coupling between disorder, rate and wettability, we define two characteristics: invasion selectivity and cooperativity.
Selectivity, evaluated here from the mean \textit{invaded} throat size $\overline{r\mathrm{_{inv}}}$, increases with $\lambda$ and $\mathrm{Ca}$, while decreasing with $\theta$ to a minimum at slow imbibition (Fig.~\ref{fig:select}a). 
Reduced selectivity implies greater dependency of the invasion pattern on the underlying pore geometry (e.g. FA and OD in Fig.~\ref{fig:patterns}).
This occurs in ordered systems ($\lambda$=0.22), where the radially-symmetric invasion (Video 1 in supplementary material) samples nearly-equally all throat sizes (retaining the triangular distribution, Fig.~\ref{fig:select}c) such that the mean invaded size is close to the mean ($\overline{r}$=0.54$a$). 
In contrast, high disorder provides constrictions with high entry pressures that remain non-invaded (Video 2 in supplementary material), leading to a skewed distribution with a larger mean $\overline{r\mathrm{_{inv}}}$. 
Such high selectivity is mostly pronounced at slow drainage, where small pores are completely avoided (Fig.~\ref{fig:select}a--b). 
In imbibition, since cooperative invasion occurs at lower capillary pressures than burst instabilities~\cite{Holtzman2015}, smaller pores can be invaded than in drainage, reducing $\overline{r\mathrm{_{inv}}}$.

\begin{figure}[h]
\centering
\includegraphics[width=.45\columnwidth]{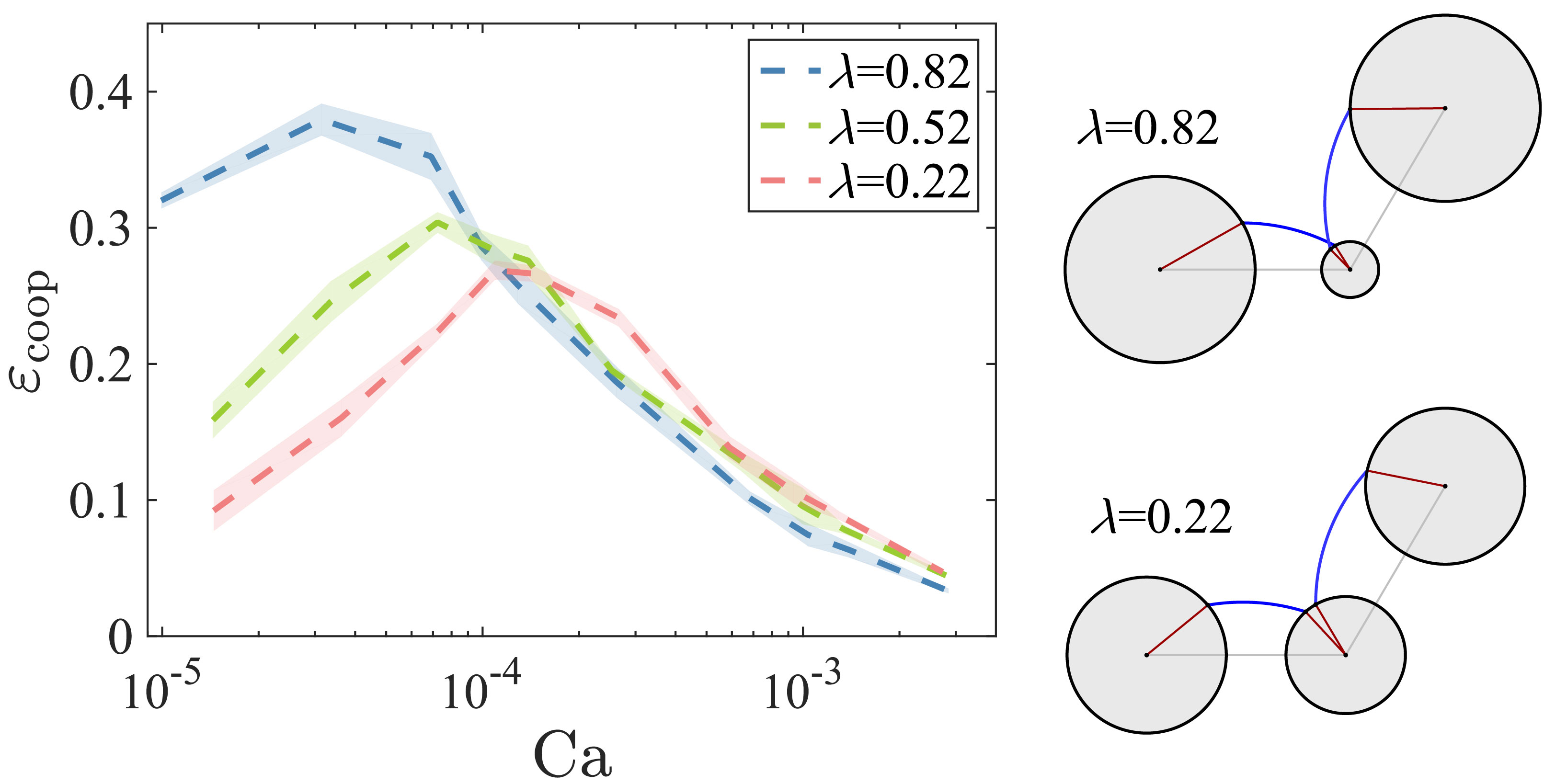}
\caption{In imbibition, a significant portion of the pores (measured by the cooperativity $\varepsilon_{coop}$) are invaded by cooperative pore filling events (``overlaps'')~\cite{Holtzman2015}. Lowering disorder hinders cooperativity, as demonstrated by the side panels showing two adjacent menisci with similar curvatures. 
At high flow rates, increasing dominance of viscous instabilities suppresses overlaps~\cite{Holtzman2015}, decreasing $\varepsilon_{coop}$. 
Lines and shading show ensemble mean and standard deviation among four realizations. 
\label{fig:coop}}
\end{figure}

Cooperativity $\varepsilon_{coop}$ is the relative number of nonlocal, cooperative pore filling events (overlaps). Overlaps, which increase in occurrence with $\theta$ (and do not appear below $\sim$60$^{\circ}$), stabilize imbibition patterns by suppressing fingering and trapping~\cite{Holtzman2015}. 
Schematic side panels in Fig.~\ref{fig:coop} show two pairs of adjacent menisci at a similar curvature for $\lambda$=0.82 and 0.22, demonstrate that pore-size homogeneity hinders overlaps. 
Since disorder also promotes trapping, the two effects counteract, leading to the small sensitivity to $\lambda$ in slow imbibition; this is particularly noticeable for $A\mathrm{_{front}}$, which is more affected by trapping than $A\mathrm{_{inter}}$ or $w$ (Fig.~\ref{fig:Linter_Lfront_w}d--f).
Increasing rates diminishes both selectivity and cooperativity, as pressure screening results in thin fingers invading only into more permeable regions~\cite{Holtzman2015} (Figs.~\ref{fig:select}--\ref{fig:coop}). 
Consequently, the mechanism for invasion selectivity changes with rates: at low $\mathrm{Ca}$ it is the high capillary thresholds along the interface (locally), whereas at high $\mathrm{Ca}$ the flow resistance--a nonlocal effect which depends on the \textit{connectivity} of larger pores--becomes dominant.

\subsection*{Trapping and sweep efficiency} 
\begin{figure}[h]
\centering
\includegraphics[width=.8\columnwidth]{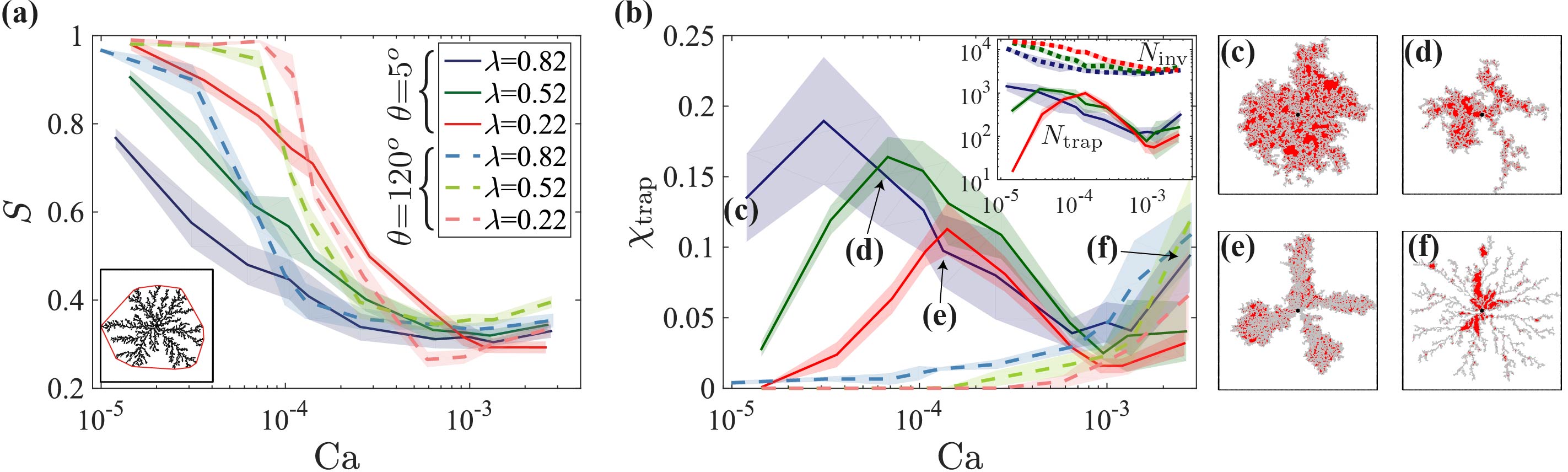}
\caption{Both sweep efficiency $S$ and trapping fraction $\chi_{\mathrm{trap}}$ depend on the flow rate $\mathrm{Ca}$, wetting angle $\theta$ and disorder $\lambda$ in a complex, non-monotonic manner. 
Sweep efficiency $S$ improves with decreasing $\lambda$ and $\mathrm{Ca}$ and increasing $\theta$ (a).  
Rate has the largest impact on sweep, as viscous instabilities leave most of the defending fluid in place, greatly reducing sweep despite the reduced trapping, $\chi_{\mathrm{trap}}$ (b). 
The trapping mechanism changes with $\mathrm{Ca}$ from multiple patches distributed \textit{within} the thick, capillary fingers (c--d) to fewer patches in between the thin viscous fingers (f); both mechanisms are suppressed by increasing $\theta$ and decreasing $\lambda$. 
Lines and shading in a--b depict the ensemble mean and standard deviation. 
Patterns in panels c--f (trapped and invaded pores in red and gray) refer to points on the curve for $\lambda=0.82$ and $\theta$=5$^{\circ}$ in b (blue solid line). 
\label{fig:sweep}}
\end{figure}

One of the most important characteristics of fluid displacement from a practical standpoint--describing the ability to produce oil or water and displace a contaminant--is the sweep efficiency.
Sweep is intimately related to trapping, leaving behind isolated patches of the defending fluid. 
%
%
We define sweep $S$ and trapping fraction $\chi_{\mathrm{trap}}$ from the following dimensionless volumetric ratios: $S = V_\mathrm{inv} / V_\mathrm{tot}$, where $V_\mathrm{tot}$ is the volume of the convex polygon wrapping the pattern (Fig.~\ref{fig:sweep}a, inset), and $\chi_{\mathrm{trap}} = V_{\mathrm{trap}} / V_{\mathrm{inv}}$, where $V_{\mathrm{trap}}$ is the volume of trapped fluid (Fig.~\ref{fig:sweep}c--f, in red).

The complex interplay between disorder, wettability and rates is demonstrated by the nonlinear, non-monotonic behavior of sweep and trapping~\cite{Liu2013}. 
Both lowering disorder $\lambda$ and increasing wetting angle $\theta$ provide a similar effect: suppressing trapping hence increasing sweep, to almost 100\% at low rates (Fig.~\ref{fig:sweep}a--b).  
Flow rate strongly impacts both $S$ and $\chi_{\mathrm{trap}}$. In slow drainage (low $\mathrm{Ca}$ and $\theta$) and high $\lambda$, the patchy CF-like pattern traps a significant portion of the defending fluid (Fig.~\ref{fig:sweep}b; see pattern in Fig.~\ref{fig:sweep}c).
At slow imbibition (high $\theta$), lowering disorder inhibits both cooperative pore filling (Fig.~\ref{fig:coop}) and trapping, effects which offset each other, explaining the observed minimal impact of $\lambda$ (cf. Fig.~\ref{fig:Linter_Lfront_w}d--f).
Increasing the rate promotes viscous instabilities and invasion by thin fingers which leave most of the defending fluid behind (Fig.~\ref{fig:sweep}f), minimizing sweep despite of the reduced trapping. 
Rates change not only the trapped volume, but also the underlying mechanism: from multiple patches \textit{within} thick fingers (CF) at low $\mathrm{Ca}$ (Fig.~\ref{fig:sweep}c--d) to fewer patches trapped \textit{between} thin fingers (VF) at high $\mathrm{Ca}$ (Fig.~\ref{fig:sweep}f).


\subsection*{Scaling analysis: Introducing disorder effects} 
\label{RescalingCa} 


In light of the complex effect of disorder, and, in particular, the ambiguity in the literature regarding its interplay with viscous forces, we address this analytically via scaling, deriving a disorder-dependent number $\mathrm{Ca}^*$. 
We begin with the classical definition, $\mathrm{Ca}=\delta p_{vis} / \delta p_{cap}$, where $\delta p_{vis}$ and $\delta p_{cap}$ are the characteristic viscous and capillary forces~\cite{lenormand90-liquids}.
The force driving the growth of individual fingers perpendicular to the interface (along the direction of the externally-applied pressure drop) $\delta p_{vis}$ is evaluated from the pressure drop in the viscous defending fluid over a characteristic length $L_{vis}$, $\delta p_{vis}\sim\nabla p_{vis}L_{vis}$. Here $\nabla p_{vis}\sim\mu_{d}v/k$ is the gradient and $k$ is the medium's permeability.
The capillary force is evaluated from a characteristic entry pressure, $\delta p_{cap} \sim \gamma/  r_c$, where $r_c$ is a characteristic aperture, providing~\cite{lovollmeheust04}
\begin{equation}
\mathrm{Ca} = \frac {  \mu_d v} {\gamma} {\ }  \frac { r_c L_{vis}} {k}.
\label{eq:Ca}
\end{equation}
The classical expression~\cite{lenormand90-liquids} $\mathrm{Ca} = \mu_{d}v/\gamma$ is obtained by evaluating $\delta p_{vis}$ at the pore-scale, $L_{vis} \sim \overline{r}$, together with the simplifying assumptions $r_c \sim \overline{r}$ and $k \sim {\overline{r}}^2$.

Here, we point to an inherent shortcoming in the above definition when disorder is introduced: the assumption $k \sim {\overline{r}}^2$ ignores the effect of interpore connectivity, which controls the resistance to viscous pressure dissipation (flow). We claim that disorder in pore throat sizes $\lambda$ affects \textit{both} (i) capillary forces $\delta p_{cap}$ through the entry threshold distribution which controls the invasion order along the interface (locally), as well as (ii) viscous forces through the permeability $k$, a {nonlocal} effect, not accounted for in previous works \cite{yortsosxu97, toussaintlovoll05, holtzmanjuanes10-fingfrac,Xu2014,Liu2015}.
To include the latter in our definition of $\mathrm{Ca}^*$, we evaluate $k$ as an \textit{effective} permeability: considering finger growth in the radial direction, we use the weighted harmonic mean of the local (interpore) permeabilities $k_i \sim {r_i}^2$ (i.e. resistors in series), $k^{-1} \sim \overline{1/{r^2}}$, providing 
\begin{equation}
\mathrm{Ca^*}=\mathrm{Ca} {\ }  {r_c} {\ } \overline{r} {\ } \overline{1/{r^2}}. 
\label{eq:rescaledCa}
\end{equation}
The main outcome of Eq.~(\ref{eq:rescaledCa}) is that since increasing disorder (the width of $r$ distribution) decreases the size of the smallest throats (``bottlenecks'') and thus the permeability $k$ (which is biased towards the \textit{smallest} $r$ values), it increases $\delta p_{vis}$. If the accompanying increase in $\delta p_{cap}$ is smaller, this would amount to increasing $\mathrm{Ca^*}$.


In our derivation, unlike the classical one presented above, the terms ${r_c}$ and $\overline{r}$ in Eq.~(\ref{eq:rescaledCa}) represent different characteristic lengths (though both at the pore-scale), hence are not equal: ${r_c}$ arises from the characteristic threshold, $\delta p_{cap} \sim \gamma /  r_c$, whereas $\overline{r}$ is our choice for the pore scale characteristic length for pressure dissipation, $L_{vis} \sim \overline{r}$. 
Furthermore, the choice of $r_c$ in Eq.~(\ref{eq:rescaledCa}) depends on wettability. 
In drainage, since invasion occurs mostly though burst instabilities~\cite{Holtzman2015} which strongly depend on throat sizes $r$, the \textit{distribution} is dominant; thus, we evaluate $r_c$ using the standard deviation, $r_c = \sigma (r)$. 
In imbibition, the increasing occurrence of overlaps with $\lambda$ (Fig.~\ref{fig:coop}) makes for a complex dependence of the characteristic invaded size $r_c$ on disorder. Since predicting this dependence is out of the scope of this paper, we simply use $r_c= \overline{r}$ for imbibition. 

%

\begin{figure}[h]
\centering
\includegraphics[width=.7\columnwidth]{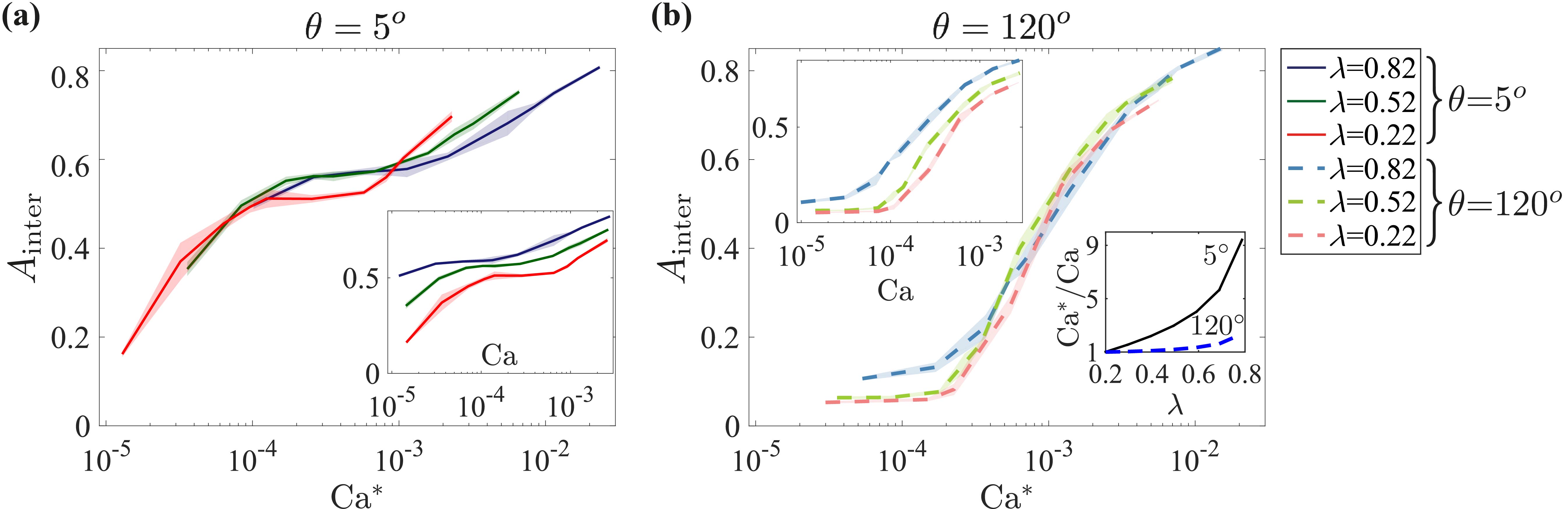}
\caption{We explain quantitatively the impact of disorder through scaling analysis. 
Plotting interfacial area $A\mathrm{_{inter}}$ against our rescaled capillary number $\mathrm{Ca}^*$ (rather than $\mathrm{Ca}$, insets) provides a partial collapse of curves of different $\lambda$. This indicates a \textit{direct} relationship between $\mathrm{Ca^*}$ and $\lambda$ (b, lower inset), suggesting a stronger impact of disorder on the \textit{viscous} forces, thus a lower transitional $\mathrm{Ca_{CF/VF}}$ at higher $\lambda$.
Our scaling also demonstrates quantitatively the diminishing effect of $\lambda$ as the wetting angle $\theta$ increases (b, lower inset; see text).
%
Lines and shading depict the ensemble mean and standard deviation. 
%
\label{fig:rescaling_Ca_Nx100}}
\end{figure}


As samples with higher $\lambda$ exhibit more tortuous, thinner fingers (at a given $\mathrm{Ca}$, see insets in Fig.~\ref{fig:rescaling_Ca_Nx100}), plotting the pattern characteristic $A\mathrm{_{inter}}$ vs. our rescaled $\mathrm{Ca^*}$ partially collapses curves of different $\lambda$, (Fig.~\ref{fig:rescaling_Ca_Nx100}; similar collapse observed for $A\mathrm{_{front}}$ and $w$). 
This indicates a \textit{direct} relationship between $\mathrm{Ca^*}$ and $\lambda$ (Fig.~\ref{fig:rescaling_Ca_Nx100}b, lower inset), namely a lower transitional $\mathrm{Ca_{CF/VF}}$ value at higher $\lambda$, in par with the inverse relationship (and increase in $\mathrm{Ca_{CF/VF}}$) in the literature \cite{yortsosxu97, toussaintlovoll05, holtzmanjuanes10-fingfrac,Xu2014,Liu2015}. 
Our scaling analysis also quantifies the decreased sensitivity of the emergent pattern to $\lambda$ as the wetting angle $\theta$ increases (highlighted in Fig.~\ref{fig:Linter_Lfront_w}d--f): $\mathrm{Ca^*} / \mathrm{Ca}$ is $\sim$10 times larger for $\lambda$=0.82 than for $\lambda$=0.22 for drainage ($\theta$=5$^{\circ}$), vs. only $\sim$2.5 for imbibition ($\theta$=120$^{\circ}$), see inset in Fig.~\ref{fig:rescaling_Ca_Nx100}b.
%
%
We note that while our scaling is a valuable quantitative explanation of our findings, the expression in Eq.~(\ref{eq:rescaledCa}) and consequent values (Fig.~\ref{fig:rescaling_Ca_Nx100}) were derived for our 2-D medium with uniform distribution of round particles; extension to other types of media is of interest for future research.


%

%

\section*{Discussion}

We study the impact of random pore size disorder on the fluid-fluid displacement in porous media. 
We use a novel pore-scale model which provides a mechanistic description of both drainage and imbibition. Our numerical results are validated against micromodel experiments, and explained theoretically by scaling analysis. 
To the best of our knowledge this is the first systematic, quantitative study of the impact of disorder and its interplay with flow rates and wettability.

We show that lowering disorder $\lambda$ increases the pattern compactness, with a smoother fluid-fluid interface. We demonstrate that the effect of disorder depends on its coupling with flow rate and wettability: (i) at low rates, a more pronounced impact of $\lambda$ on capillary thresholds and hence on the order of invasion and trapping; (ii) at high rates, increasing $\lambda$ strongly increases the viscous pressure drop by decreasing pore connectivity, promoting pressure screening and viscous instabilities, increasing the number of thin, tortuous fingers; and (iii) increasing the wetting angle $\theta$ stabilizes the invasion pattern by reducing trapping and smoothing the interface, thus minimizing the impact of disorder. 
%
%
At low $\lambda$, the underlying pore geometry (lattice structure) becomes dominant, in particular in imbibition, leading to symmetrical growth of faceted, compact pattern at low $\mathrm{Ca}$ and dendritic fingers at high $\mathrm{Ca}$.

While our observation that decreasing disorder increases the sweep efficiency agrees with previous studies \cite{chenwilkinson85,yortsosxu97,Cieplak1990,MartysPRB1991, toussaintlovoll05, holtzmanjuanes10-fingfrac, Xu2014, Liu2015}, our interpretation does not: we emphasize the previously ignored impact of pore-scale disorder on the effective, sample-scale connectivity (permeability), and thus on the viscous forces. 
We quantify this effect by deriving a disorder-dependent capillary number $\mathrm{Ca}^*$ providing a \textit{direct} relationship between $\mathrm{Ca^*}$ and $\lambda$, that is a lower transitional $\mathrm{Ca_{CF/VF}}$ value at higher $\lambda$, in par with previous works \cite{yortsosxu97, toussaintlovoll05, holtzmanjuanes10-fingfrac,Xu2014,Liu2015}. 
Our findings delineate the complex interplay among disorder, rates and wettability, via a simplified 2-D model. Valuable future work includes generalization of the model towards 3-D granular pack with correlated heterogeneities (more representative of soils and rocks), inclusion of additional mechanisms such as droplet fragmentation or film flow, and a more rigorous experimental validation by improving manufacturing accuracy and quantitative analysis. 


In summary, our work offers a fresh perspective into a long-standing paradigm of significance in processes where efficiency of fluid extraction/injection and mixing and reaction of the fluids are crucial. Increasing disorder promotes fingering, which limits the recovery of subsurface fluids such as hydrocarbons~\cite{LakeOil2014} or contaminants~\cite{Nadim2000}, while increasing interfacial areas which controls fluid mixing and reactions~\cite{Jha2011} in microfluidics~\cite{Stone_microfluidics2004,Kim2015}, subsurface remediation~\cite{Nadim2000}, soil aeration~\cite{Li_AerationTomato_2015}, and long-term CO$_2$ geosequestration by carbonate mineralization~\cite{Matter2016}.

\section*{Methods}

\subsection*{Model medium: 2-D analog} 

The main objective of this work is to gain \textit{fundamental} understanding of the complex interactions among disorder, wettability and rates. Therefore, we choose the simplest representation (for both numerics and experiments) of a partially-wettable, disordered porous medium that retains the most crucial aspects of the behavior of such medium. 
We consider here the following model system: cylindrical solid pillars placed on a regular lattice (Fig.~\ref{fig:model_and_expt_schematics}), noting that consideration of the convoluted 3-D pore geometry (relevant for instance to natural sediments) would hinder basic insights, and thus should be left to a later stage. 
We introduce disorder through variations in particle sizes, rather than through randomness in particle positions or in the coordination number of pores, since the former allows a more rigorous quantitative comparison among different samples. 
Similarly, our choice of triangular lattice limits the number of menisci which can invade into a pore simultaneously to two, retaining the crucial aspect of cooperative pore filling while avoiding the ambiguity of resolving a burst on a square lattice~\cite{Cieplak1990}. 
%


%
We construct square-shaped samples of size $L=N_x a$ by placing $N_x \times N_y$ particles ($N_y \times 2N_x$ pores), where $N_y$=$2N_x/\sqrt 3$ and $a$ is the lattice spacing. Each particle triplet defines a pore with volume $V$ connected to three neighboring pores by throats with aperture $0 < r\lesssim a$; to avoid closed throats due to particle overlap we enforce $\bar{d}< a/(1+\lambda)$, see Fig.~\ref{fig:model_and_expt_schematics}c.
Pillar diameters are much smaller than their height, $d\ll h$,
such that the in-plane curvature is much larger than the out-of-plane curvature, $R_{in}\ll R_{out}$. This allows us to approximate the meniscus as 2-D, with a radius of curvature $R=1/\left(R_{in}^{-1}+R_{out}^{-1}\right)\approx R_{in}$.

%


\subsection*{Numerical simulations}



Simulations are conducted with the model of Holtzman and Segre~\cite{Holtzman2015}, summarized below (see supplementary material for further details). This model can accommodate a wide range of properties and conditions including flow rates, wetting properties and viscosities. To the best of our knowledge, it is the only pore-scale model which provides a mechanistic description of the dynamics of partially-wetting invasion in a large, disordered domain. Previous models were limited by computational costs~\cite{Meakin2009,Joekar-Niasar_CREST_2012,Bultreys2016}, compromising the dynamics by assuming quasi-static displacement~\cite{Cieplak1988,Cieplak1990} or the size and heterogeneity of the domain~\cite{Liu2013,Liu2015}.

We represent the fluid-fluid interface as a sequence of menisci, where each meniscus is a circular arcs which intersects a pair of particles at the prescribed contact angle $\theta$, and has a radius of curvature $R=\gamma / \Delta p$ according to the Young-Laplace law. Here $\Delta p$ is the pressure jump across the meniscus.
The angle $\theta$ is measured through the defending fluid (i.e. $\theta< 90^{\circ}$ for drainage), and represents an \textit{effective advancing} angle, which includes dynamic effects~\cite{Meakin2009}.
Knowledge of $R$ and $\theta$ allows to resolve analytically the position and hence stability of each meniscus.  
%
%

A crucial feature of our model is its account for cooperative pore-filling (overlaps~\cite{Cieplak1990}), which become dominant at high $\theta$. 
%
Overlap is a nonlocal mechanism as it is affected by menisci in multiple pores, acting to smooth the interface~\cite{Cieplak1990,Holtzman2015}. 
Menisci are tested for three types of capillary instabilities~\cite{Cieplak1990}: (1)
Haines jump or \emph{burst}, when the curvature exceeds a threshold set by the local geometry;
(2) \emph{touch}, when a meniscus intersects a third, downstream particle; and
(3) \emph{overlap} of adjacent menisci, destabilizing each other [Fig.~\ref{fig:model_and_expt_schematics}(c--e)]. 
Once destabilized, the meniscus incipiently invades the downstream pore, at a rate evaluated from the fluid's viscous resistance (see supplementary material for details).
The finite pore filling time in our model captures the invasion dynamics, including the spatiotemporal nonlocality associated with rapid interfacial jumps and the consequent interface readjustments and pressure screening~\cite{lovollmeheust04,Armstrong2013}, overcoming a long-standing computational challenge~\cite{Meakin2009,Joekar-Niasar_CREST_2012}.

\subsection*{Experimental setup}
\label{expt_setup}

Micromodels allow visualization of the solids and fluids, making them a highly useful tool to study fluid displacement.
Here, we use 3-D printing (``stereolithography''), an additive method fabricating free-form solids using a computer-aided design (CAD) file. 
A ProJet 3500 Multi Jet Printer (3D Systems Inc.) with ``VisiJet M3 Crystal'' plastic provided in-plane and out-of-plane resolution of 50 and 32 $\mu${m}, respectively \cite{snyder20143d}. The limited resolution resulted in rough, non-round pillars, and, though rare, a few that are misplaced (off-lattice), see insets in Fig.~\ref{fig:model_and_expt_schematics}f.
We fabricated two cells of geometry similar to the numerical samples ($L=100 a$) with $\lambda=0.22$ and $0.52$, however with lattice spacing $a=1$ {mm} (vs. 0.5 mm in simulations).
%
Pillar height of $h=3$ {mm} was chosen to ensure sufficiently large aspect ratio $h/ r$ such that the meniscus can be approximated as 2-D, while small enough to minimize hydrostatic pressure and thus 3-D, out-of-plane effects inducing vertical flows. 
The lattice is embedded in a frame with a trench (of width 1 {cm}) around it to ensure evenly distributed withdrawal. 
The chip is sandwiched between two 1 cm-thick acrylic plates, with an outlet cross-sectional area $A_{out} \approx$ 5.5$\cdot$10$^{-4}$ m$^{2}$.


The cell is initially filled with glycerol ($\sim$99.5\% purity with $\mu_d \approx 1$Pa$\cdot$s and $\gamma \approx 6.3 \cdot 10^{-2}$N/m at $\sim$25$^{\circ}$C, Gadot Chemical Terminals LTD.). 
A high-viscosity fluid was selected to probe high $\mathrm{Ca}$, where the CF--VF transition occurs. 
%
The contact angle (for both acrylic and VisiJet M3 Crystal) was measured to be $75^{\circ}\pm5^{\circ}$ using a sessile drop goniometer (EasyDrop DSA20E, Kr\"uss GmbH).
%
Glycerol is withdrawn from ports at the cell corners by a syringe pump (Harvard Apparatus PHD ULTRA 70-3007) at a fixed rate $Q$, letting air ($\mu_i = 1.8 \cdot 10^{-5}$Pa$\cdot$s) invade freely through a central inlet (Fig.~\ref{fig:model_and_expt_schematics}f). 
$\mathrm{Ca}$ values in Fig.~\ref{fig:experiments} correspond to $Q$=0.2, 0.5, 2, 5, 20, 50 and 100 ml/min.
%
The cell is held horizontally to avoid gravity effects. 
%
The pattern evolution is tracked by time-lapse photography using a CMOS camera (uEye 3520, IDS 1.3MP). We illuminate the cell from above with LEDs. 
To improve image contrast, the bottom acrylic plate was painted black, and the inner side of the top plate was sandpapered, causing dry areas to appear opaque.


%


\section*{Acknowledgements}

The author thanks E. Segre and A. Naftali for invaluable assistance with the simulations (E.S. and A.N.) and experiments (E.S.). Financial support by the Israeli Science
Foundation (\#ISF-867/13), the United States-Israel Binational Science
Foundation (\#BSF-2012140) and the Israel Ministry of Agriculture
and Rural Development (\#821-0137-13) is gratefully acknowledged.

\section*{Author contributions statement}
R.H. conceived the study, acquired the data, analyzed the results and wrote the manuscript. 

\section*{Additional information}

\noindent\textbf{Supplementary information} containing (i) details of the numerical model; (ii) simulation parameters; and (iii) experimental videos showing the pattern growth dynamics accompanies this paper at http://www.nature.com/srep.

\noindent\textbf{Competing financial interests}: The author declares no competing financial interests.

\noindent\textbf{How to cite this article: }



\end{document}